\documentclass[rnote]{aa}

\usepackage{graphics}
\usepackage[varg]{txfonts}
\usepackage{natbib}
\bibpunct{(}{)}{;}{a}{}{,}
\usepackage{color}

\begin{document}

\title{Modeling non-thermal emission from stellar bow shocks}

\author{V. Pereira\inst{1}
  \and J. L\'opez-Santiago\inst{1}
     \and M. Miceli\inst{2,3}
     \and R. Bonito\inst{2,3}
     \and E. de Castro\inst{1}}


\institute{Dpto. de Astrof\'isica y CC de la Atm\'osfera, Universidad Complutense de Madrid, E-28040 Madrid, Spain
  \and INAF-Osservatorio Astronomico di Palermo, Piazza del Parlamento 1, I-90134 Palermo, Italy
  \and Dipartimento di Fisica e Chimica Universit\`a di Palermo, Piazza del Parlamento, I-90134 Palermo, Italy}

\date{Received 16 December 2015/ Accepted 16 February 2016}


\abstract {Runaway O- and early B-type stars passing throughout the interstellar medium at supersonic velocities and characterized by strong stellar winds may produce bow shocks that can serve as particle acceleration sites. Previous theoretical models predict the production of high energy photons by non-thermal radiative processes, but their efficiency is still debated.} {We aim to test and explain the possibility of emission from the bow shocks formed by runaway stars traveling through the interstellar medium by using previous theoretical models.}  {We apply our model to AE Aurigae, the first reported star with an X-ray detected bow shock, to BD+43 3654, in which the observations failed in detecting high energy emission, and to the transition phase of a supergiant star in the late stages of its life.} {From our analysis, we confirm that the X-ray emission from the bow shock produced by AE Aurigae can be explained by inverse Compton processes involving the infrared photons of the heated dust. We also predict low high energy flux emission from the bow shock produced by BD+43 3654, and the possibility of high energy emission from the bow shock formed by a supergiant star during the transition phase from blue to red supergiant.} {Bow shock formed by different type of runaway stars are revealed as a new possible source of high energy photons in our neighbourhood.}

\maketitle


\section{Introduction}
Runaway stars are stars that exceed the typical velocities of the stars in the galaxy and can reach more than 200 km s$^{-1}$ \citep[e.g.,][]{bla61, gyb86}. To distinguish them from the stars with typical velocities and dispersions of the order of 10 km s$^{-1}$, runaway stars are considered those with $V_*>$ 30 km s$^{-1}$. When these stars move supersonically through the interstellar medium (ISM) and the medium density is dense enough, the stellar wind sweeps and piles up the circumstellar dust, leading to the formation of a bow shock in the direction of their movement. 

Relativistic particles can be accelerated by strong shocks, and the interaction of these particles with the matter, the radiation, and the magnetic field of the bow shock region can produce non-thermal emission. \citet{ben10} detected radio emission from the bow shock produced by the runaway star BD+43 3654, consistent with synchrotron radiation produced by the interaction of the relativistic electrons with the magnetic field of the acceleration region. Later, \citet{vyr12} presented a non-thermal emission model that predicted the production of high energy photons in these acceleration regions in a number sufficiently large be detected in X-rays. \citet{lop12} published the first detection of X-ray emission from a bow shock produced by a runaway star (AE Aurigae), whilst \citet{ter12} reported the non-detection of X-ray emission in the bow shock region formed by BD+43 3654 from a long time X-ray observation with \textit{Suzaku}.

Due to their strong winds, O and early-B runaway main sequence type stars have been traditionally considered to be the best candidates to form strong bow shocks. With masses higher than 10 $M_\odot$ and normalsizes between 10 and 50 kK, in their short life ($\le$20 Myr) these stars loss mass at a rate of $\dot{M} \sim 10^{-7} - 10^{-5}$ $M_\odot$  yr$^{-1}$ and their stellar winds, with velocities ranging from 1000 to 3000 km $s^{-1}$, transfer great amounts of mechanical energy to the circumstellar medium. Last searches of runaway O and early-B are those of \citet{gyb86}, \citet{gie87}, \citet{mof98, mof99} and \citet{mai04}. More recently, an extensive kinematic and probabilistic study has led to the identification of 2500 candidates of  runaway stars from the data of the \textit{Hipparcos} catalog \citep{tet10}. 

In this work we aim to explain the X-ray emission from the bow shock formed by AE Aurigae and the non X-ray detection from the bow shock of BD+43 3654, and also to extend the study to other runaway stars interacting with the ISM. For this purpose, we follow the model derived by \citet{vyr12}, where different non-thermal processes are proposed to explain the origin of the emission in these shocks. Further details on the non-thermal processes are also found in \citet{ada80}, \citet{kel06}, \citet{lon11} or \citet{ryl79}, and other contributions to the explanation of the high energy emission from the bow shocks produced by runaway stars are those of \citet{ben10}, \citet{ter12}, and \citet{vyr14}.

 \begin{figure}
      \resizebox{\hsize}{!}{\includegraphics{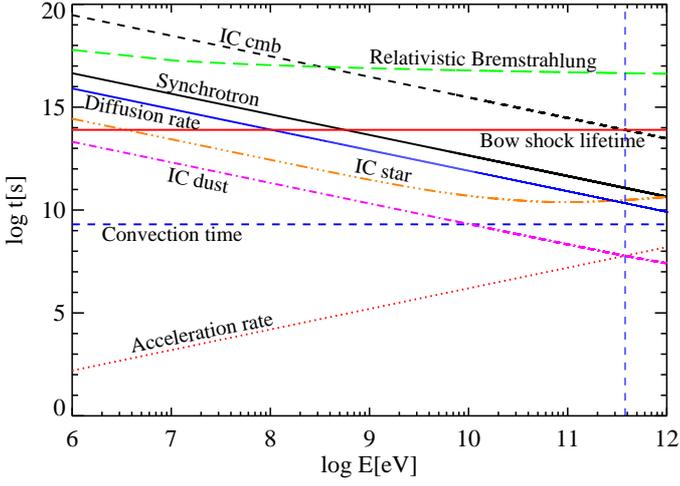}}
       \caption{Leptonic cooling time rates, acceleration rate, convection time, diffusion rate and bow shock lifetime for the parameters of the bow shock formed by AE Aurigae (see Sec. \ref{sec:AE}). The vertical dashed line indicates the maximum energy.}
     \label{fig:AEtimes}
\end{figure}

 \begin{figure}
      \resizebox{\hsize}{!}{\includegraphics{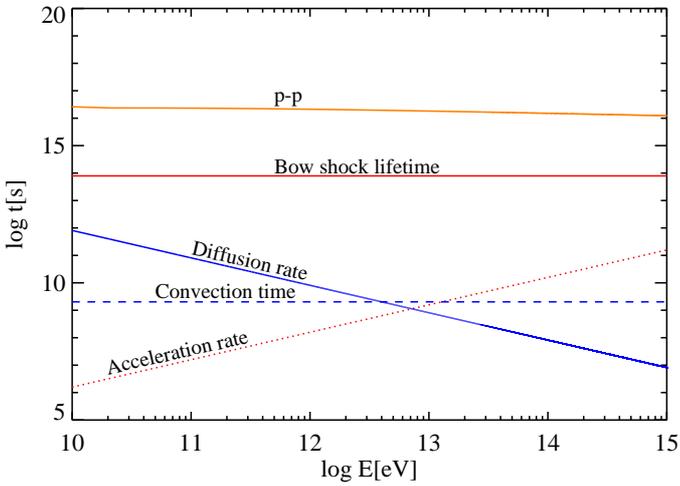}}
       \caption{Proton-proton inelastic collisions time rate, convention time, diffusion rate, and bow shock lifetime for the parameters of the bow shock formed by AE Aurigae (see Sec. \ref{sec:AE}). The cooling time rates represent the time a particle with energy E needs to cool completely by a given process.}
     \label{fig:timeprotons}
\end{figure}


\section{Applications}
\label{sec:applications}

\citet{lop12} reported the first X-ray detection of a bow shock formed by a runaway star (AE Aurigae). \citet{ter12} did not succeed in detecting significant diffuse X-rays from the bow shock formed by the runaway star BD+43 3654. In this section, we apply our non-thermal model to clarify the origin of the X-ray emission detected from the bow shock formed by the star AE Aurigae (HIP 24575), and to try to explain the non detection in this band of the bow shock formed by BD+43 3654. Later, we extend our study to other stars to test wether the bow shocks they form can also be a source of high energy photons in our galaxy. In particular, we study the bow shock formed by a supergiant star during the transition from blue to red using the star Betelgeuse as a test of our computations.  

The power available in the acceleration region of a bow shock is given by the volume ratio, i.e., $L = V L_T / V_{R_0}$, where $L_T$ is the kinetic energy released by the stellar wind per second, $V_{R_0}$ is the volume of a sphere of radius $R_0$, and $V$ is the volume of the acceleration region. $R_0$ is the so called standoff radius expressed by \citep{wil96}
\begin{equation}R_0=\sqrt{\frac{\dot{M} V_w}{4\pi\rho_{ISM} V_{\star}^2}}.
\label{eq:wilkin}
\end{equation}
Here, $\dot{M}$ is the mass-loss rate, $V_w$ the wind velocity, $\rho_{ISM}$ ($\rho_a=n_H\mu$, where $n_H$ is the ambient column density in cm$^{-3}$ and $\mu=2.3\times10^{-24}$ g the mass per H atom) is the ambient density and $V_{\star}$ is the stellar velocity. We assume that only a small fraction $q_{rel}$ of $L$ is turned into relativistic particles and thus the total energy of the accelerated charged particles is $L_{rel} = q_{rel} L$. These relativistic particles can be either electrons or protons, so $L_{rel}$ has both a leptonic and an hadronic component, $L_{rel}=L_p + L_e = aL_e + L_e$  where $a$ is the ratio of relativistic protons to electrons. 

In the three simulations we perform we assume the medium the stars are traveling through to be homogeneous and free of clumps and that the IC seed fields are isotropic. In Table \ref{table:1} we list the stellar parameters of these three stars and the free parameters we used in our model. The electron to proton ratio, although a free parameter of the model, was fixed to $a = 1$ (see explanation in Sec. \ref{sec:AE}). Both the electron injection index and the fraction of relativistic energy were fitted to the available observational data points. The IR shock widths were taken from \citet{per12}. However, the non-thermal emission is produced from a smaller region, concentrated in a small knot at the bow shock apex, and hence the final shock width is also obtained from fitting the observational data.

\begin{table*}
\caption{Model parameters}
\label{table:1} 
\centering 
\begin{tabular}{lllll} 
\hline\hline 
\multicolumn{5}{c}{Stellar parameters}                                          \\
\hline
\multicolumn{2}{c}{Parameters}                                                                  &AE Aurigae                         &BD+43 3654                     &Betelgeuse                                                                     \\
\hline 
$\dot{M}$       &Wind mass-loss rate    [$M_{\odot}$ yr$^{-1}$]                         &10$^{-7}$                              &10$^{-5}$                      &3.2 $\times$ 10$^{-7}$                                                      \\
$V_w$           &Wind velocity [km s$^{-1}$]                                                    &1500                                         &2300                           &270                                                                            \\
$R_0$           &Standoff radius [pc]                                                                   &0.082                                         &1.5                                    &0.63                                                                                 \\
$L_{\star}$     &Luminosity of the star [$ L_{\odot}$]                                  &0.7 $\times$ 10$^{5} $              &8 $\times$ 10$^{5}$    &4.8 $\times$ 10$^{4}$                                                       \\
$T_{\star}$     &Stellar normalsize [kK]                                                               &32                                                 &39                                     &9                                                                                         \\
$R_{\star}$     &Stellar radius [$R_{\odot}$]                                                    &8.9                                            &19.4                                 &90                                                                                     \\
$n_H$           &Ambient medium density [cm$^{-3}$]                                     &2.3                                                 &5                                      &1.3                                                                                         \\
$V_{\star}$     &Stellar velocity [km s$^{-1}$]                                                 &150                                         &66                                     &50                                                                                         \\
$d$                     &Distance [pc]                                                                          &550                                         &1450                           &200                                                                                 \\
Age                     &Bow shock lifetime [Myear]                                                     &2.5                                                 &1.6                                    &30                                                                                         \\              
\hline  
\multicolumn{5}{c}{Free parameters in the model}                                                                                                                                                                                                                                \\
\hline
$\Delta$        &Shock width [$R_0$]                                                                    &0.3                                                 &0.08                           &0.25                                                                                 \\
$\alpha$                &Electron injection index                                                               &2                                              &2.1                                    &2                                                                                      \\
$q_{rel}$       &Fraction of relativistic energy                                                                &0.15                                   &0.1                                    &0.1                                                                                    \\
a                       &Electron to proton ratio                                                               &1                                              &1                                      &1                                                                                      \\
\hline 
\end{tabular}
\end{table*}

\subsection{AE Aurigae (AE Aur)}
\label{sec:AE}

The runaway star AE Aur was ejected from the Orion nebula cluster after an encounter between two, systems about 2.5 million years ago, and is considered a runaway star for its high velocity, $v_* \approx$ 150 km s$^{-1}$ \citep{per12}. As a result of this interaction, AE Aur and $\mu$ Col (both O9.5 type stars) were expelled at great velocities, while $\iota$ Ori remained forming an eccentric binary system with the two more massive stars \citep{hoo00}. In its path, AE Aur encountered the dense molecular cloud IC 405, with a density $n_H \sim$ 3 cm$^{-3}$ \citep[see][]{per12}. The interaction between the strong stellar wind and the dust resulted in the formation of a bow shock, which was first detected in the mid-IR by \citet{vym88} using IRAS. The terminal wind velocity of the star is $v_{w} \approx$ 1500 km s$^{-1}$ \citep{hub11} and its mass-loss rate $\dot{M} \approx$ 10$^{-7} M_{\odot}$ yr$^{-1}$ \citep{ful06}.
The cooling time rates obtained for these parameters were represented in Fig. \ref{fig:AEtimes}.

\begin{figure}[t]
     \resizebox{\hsize}{!}{\includegraphics{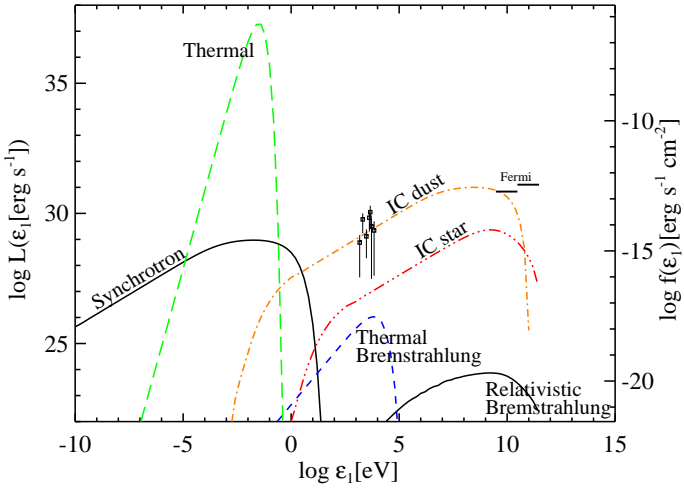}}
       \caption{Synthesized luminosity spectrum for the bow shock formed by the star AE Aur. The \textit{Fermi}  limit of detection
from \citet{ace15} and the XMM-\textit{Newton} data points are also included.}
     \label{fig:AElum}
\end{figure}

In Fig. \ref{fig:AElum} is represented the synthesized luminosity spectrum of the bow shock formed by the star AE Aur. The X-ray emission from the bow shock reported by \citet{lop12} is concentrated in a small knot north east of the star and can be explained by the IC emission of the heated dust, with a very small contribution of the IC of the starlight. The synchrotron emission is important at radio frequencies, and the IR is dominated by the thermal emission. The thermal and the relativistic Bremstrahlung are negligible compared to the IC of the dust and to the IC of the starlight photons. The IC of the CMB results very faint. The XMM-\textit{Newton} EPIC pn data obtained by \citet{lop12} are included, and also the \textit{Fermi}  limit of detection from \citet{ace15} is shown.  

The hadronic emission produced by the proton-proton inelastic collisions is negligible, revealing that the energy released during the shock interaction is too low to produce high energy hadronic emission.

On the other hand, if we consider a = 100, as in the cosmic rays case \citep{gin64}, we can not reproduce the X-ray emission detected. In Fig. \ref{fig:timeprotons} we proved that there are not enough p-p collisions and hence there is no generation of secondary electrons that can produce the required IC flux to explain the X-ray emission, whilst, as for a = 100 the energy that goes into electrons is dramatically reduced, neither the primary electrons are able to reproduce the X-ray flux observed. 

\subsection{BD+43 3654}

\citet{cyp07} concluded that the origin of the high velocity of the star BD+34 3654 was the dynamical interaction between two massive binaries. On the contrary, \citet{gyb08} determined that this star was formed from the stellar fusion of two stars after the encounter of two binary systems, originally with $\sim$ 35 and $\sim$ 50 - 60 $M_{\odot}$ , respectively. The high linear momentum of the star indicates that the encounter was very energetic. The most massive object is now seen as an O4If. 

BD+43 3654 has been reported to have formed a bow shock detected in the IR \citep{cyp07} and in radio \citep{ben10}. Both observations are coincident and extensive.

Assuming the velocity of this star is 67 km s$^{-1}$ \citep{kob10}, the mass-loss rate 10$^{-5}$ $M_{\odot}$ yr$^{-1}$ \citep{mar04, rep04}, the stellar wind velocity 2300 km s$^{-1}$ \citep{how97}, and that the standoff radius is  $R_0$ = 1.5 pc \citep{per12}, from Eq. 1 we derive a medium density of 5 cm$^{-3}$. The extensive emission in the radio and IR wavelengths are found at a distance consistent  with $R_0$.

BD+43 3654 has one of the greatest mass-loss rates among the O and B type stars \citep[see][]{vin01}, and the same happens with the stellar wind \citep[see][]{how97}. This great amount of released energy and a moderate velocity allow the bow shock to be formed far from the star, although the energy is so high that the star could have crossed the ISM without forming a strong shock. However, the radio and IR detections discard this possibility. In this case, the high density of the interstellar region crossed by the star favors the formation of the shock. Taking into account these considerations about the stellar parameters of this star, the values in the nominator of Eq. \ref{eq:wilkin} are revealed to be among the highest of all the stars known. On the contrary, the value of the velocity of this star, in the denominator, is moderated. Under these circumstances, a higher velocity would lead to a closer shock, whereas lower velocities would make BD+43 3654 not to be a runaway star. With such parameters, the bow shock formed by this star possibly represents an upper limit for the standoff radius of a bow shock formed by a runaway star. The radio and IR points allowed us to fit the spectrum and the values of the free parameters.

\begin{figure}[h]
     \resizebox{\hsize}{!}{\includegraphics{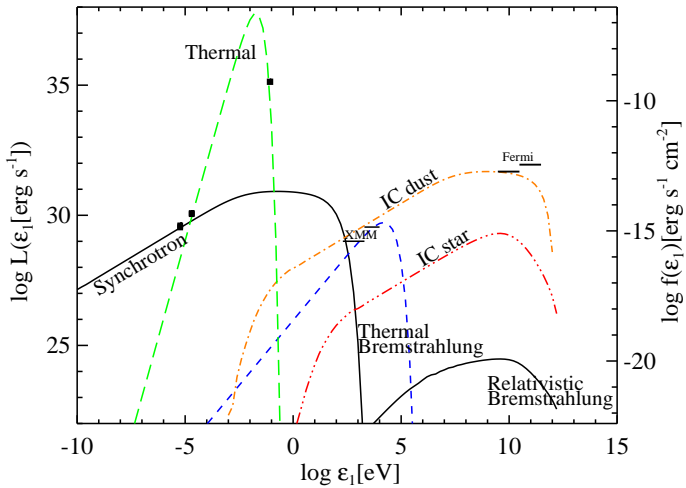}}
       \caption{Luminosity distribution for the bow shock formed by BD+43 3654. We include the MSK (at about 0.1eV) and VLA observations from \citet{ben10}. }
     \label{fig:lumBD}
\end{figure}

The spectrum of the bow shock derived from our model is shown is Fig. \ref{fig:lumBD}. \citet{ben10} suggest that the detection of VLA radio emission is associated to the cooling of electrons by synchrotron emission. However, from our computations the radio points can be explained by the sum of the non-thermal synchrotron emission and the thermal emission of the shocked dust, also consistent with the MSX data at 0.1 eV. 
Here the IC of the dust photons, plus the thermal Bremstrahlung, hardly reach the XMM-\textit{Newton} detection limit proposed by \citet{has01} for 100 ks. This suggests that the bow shock of BD+43 3654 could be detectable through longer observations. Up to now, BD+34 3654 was observed with XMM during 47 ks, which is not enough to detect the bow shock X-ray emission, according to our simulations. \citet{ter12} failed in detecting this bow shock with \textit{Suzaku}. From their observation, they obtained an X-ray luminosity lower limit at 1.1 $\times$ 10$^{32}$ erg s$^{-1}$, well over our predictions.

\subsection{Betelgeuse}

\citet{mac12} developed stellar evolution models and incorporated the evolving stellar wind into hydrodynamic simulations to simulate the transition of Betelgeuse from the blue supergiant (BSG) phase to the red supergiant (RSG) phase at the late stages of its life. At these phases, massive stars may undergo rapid transitions from red to blue supergiants and vice versa in the Hertzprung-Russell diagram, and the stellar wind velocities change rapidly. \citet{moh12} found that the bow shock formed by Betelgeuse is still young ($<$ 30 kyr). The total kinetic power is rather low, but the star is relatively close and a possible low X-ray luminosity can still be detectable. 

\begin{figure}[h]
      \resizebox{\hsize}{!}{\includegraphics{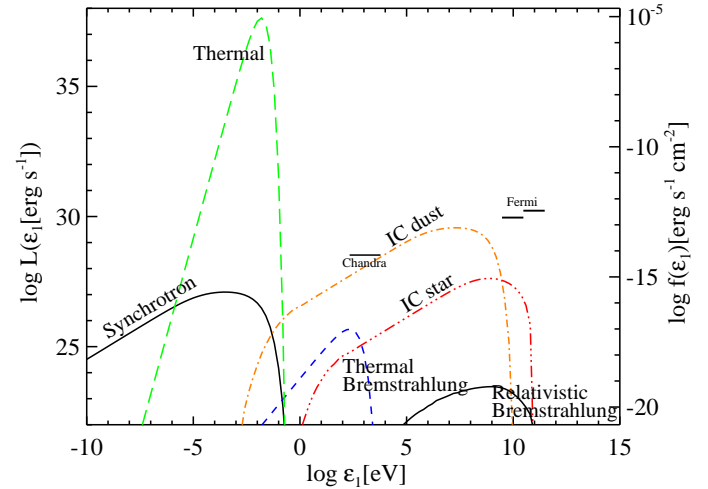}}
       \caption{Luminosity distribution for the bow shock formed by Betelgeuse during the transition phase form blue to red supergiant, previous to the current red supergiant phase.}
     \label{fig:Betellum}
\end{figure}

During the transient phase from BSG to RSG the defunct wind of the BSG still reaches a few hundreds kilometers per second. At the RSG phase the wind velocity decays down to $\sim$ 20 km/s, and hence the acceleration of particles is inefficient and the non-thermal processes are not able to produce high energy photons in a number sufficiently large to be detected with the current X-ray telescopes. Here we use the parameters derived by \citet{mac12} in their simulations to test the possibility of a non-thermal high energy emission in the transition phase from BSG to RSG, where the kinetic energy of the wind is still high. The result is showed in Fig. \ref{fig:Betellum}. The IC of the dust photons can lead to the emission of detectable X-ray photons. We also include the \textit{Chandra} detection mean limits reported by \citet{pos06} in the 0.1 - 6 keV range for this star, although less sensitive than the \textit{XMM-Newton} limit reported by \citet{has01}. Our simulation shows that the most energetic photons could be detected. However, we have to remind that the current observations of Betelgeuse do not correspond to the transition phase from BSG to RSG, but to the present RSG phase. Hence no detection is expected in the currently phase.

Although Betelgeuse is now a RSG star, we use it as a base to simulate the supergiant stars case and to test the possibility of non-thermal emission from runaway stars with its characteristics during the transition phase from BSG to RSG. The X-ray luminosity is somehow faint, but they can make up a new source of high energy photons in the neighbourhood of the Earth, although bow shocks formed during the BSG phase could be detected in X-rays at greater distances.

\section{Discussion and conclusions}

We have studied the non-thermal processes that can explain the emission detected from the bow shock formed by the star AE Aur and test the possibility of other bow shocks to emit high energy non-thermal photons in a number sufficiently large to be detected. However, the detection of X-ray emission from bow shocks around runaway stars is not expected to be found around most of them since many scenarios make the formation and detection of the bow shock unlikely. First, the runaway star must move supersonically through the ISM, whilst low density mediums do not favor the formation of bow shocks. Second, many runaway stars can form close bow shocks, which can be detectable in the IR, but the resolution of the X-rays telescopes currently available fail in resolving bow shocks with short standoff radius from their star. Furthermore, the angle formed by the line of sight and the star-bow shock line is of the utmost importance since the bow shock can have been formed far enough from the star but the angle can make it to be hidden by the runaway star or make it unresolvable from the star. On the other hand, X-ray bow shocks reach luminosities orders of magnitude fainter than the runaway stars that originate them, and hence very long exposures are needed to detect them.  

High mass-loss rates and large wind velocities favor the formation of strong shocks that can emit high energy non-thermal photons. However, they also make the bow shocks to be formed farther from the star. The available energy in the acceleration region depends on the relation between the volume of the acceleration region and the volume of a sphere of radius $R_0$. Since the acceleration region is a nearly flat region, the available kinetic energy in the acceleration region depends on the distance between the star and the acceleration region, which is equivalent to the standoff radius defined by Eq. \ref{eq:wilkin}. The volume of a sphere grows faster with the radius than the volume of the acceleration region, and thus the available energy to accelerate particles falls quickly with the standoff radius. Hence, the energy density is proportional to $R_0^{-3}$, and from Eq. \ref{eq:wilkin} the standoff radius is proportional to the inverse of the velocity of the star. These dependencies imply that if the velocity of the star were a  20\% lower , the energy density available to accelerate particles would be reduced by a 50\%. Large stellar velocities favor the formation of the bow shock at distances closer to the star, as well as high ambient medium densities. Fast winds and high mass-loss rates release great amounts of energy, which allow shocks to be formed at greater distances from the star. However, if these two quantities are very high but are not balanced by a dense medium or by a high velocity of the star, the star can cross the ISM without forming a bow shock ($R_0 \sim \infty$). This emphasize the point  that the conditions under a bow shock can be formed by a runaway star, or emit high energy photons, are very constrained, and suggest that only a small fraction of the bow shocks are expected to be detected in the X-ray band. In our three simulations the integrated X-ray flux in the 0.3 - 10 keV band is lower than 10$^{30}$ erg s$^{-1}$ (see Figs. \ref{fig:AElum}, \ref{fig:lumBD} and \ref{fig:Betellum}). These low luminosities constrain the search of bright bow shocks to stars closer than 1 kpc, due to the limited sensitivity of the X-ray telescopes and the exposure times required. Farther stars, as the case of BD+43 3654, demand very long exposures, but as the emitted X-ray flux from the bow shock is rather low, the background contamination may difficult its detection.

Runaway stars can produce bow shocks when passing through sufficiently dense clouds. These bow shocks can serve as particles acceleration sites where charged particles can cool by non-thermal processes and emit high energy photons. However, the production of these photons in a sufficiently large number to be detected is not possible in all bow shocks. Runaway O and B-type stars moving through dense molecular clouds, with the highest mass-loss rates among all the stars and the largest wind velocities, are considered the most suitable candidates to have formed a high energy emitting bow shock. However, here we have showed that runaway supergiant stars in their transition phase from blue supergiant to red supergiant also undergo important losses of mass and have moderate terminal velocities, which make them a new possible source of X-ray photons. 

From all the bow shocks detected around runaway stars, only those formed by the stars AE Aur and BD+43 3654 have been directly observed in X-rays (although the bow shock formed by BD+43 3654 has not been detected). For the case of AE Au we succeed in reproducing the non-thermal emission from the shock. For the bow shock formed by BD+43 3654 we also obtain the luminosity distribution and show that it emits high energy photons, but the fluxes are too low to be detected. Farther observations are needed to constrain the stellar parameters that favor the shock formation and that can lead to the production of enough high energy photons to be detectable in the X-ray domain, but our model sets a theoretical scenario to constrain the most reliable candidates to produce high energy emission and to test its detectability.

Here we have shown that bow shocks formed by different type of runaway stars can account for a significant fraction of the high energy photons produced in our galaxy. More energetic events in our galaxy, such as supernova remnants, produce a larger amount of high energy photons, and those sources can be detected at larger distances. However, the IC losses undergone by the electrons accelerated in the bow shock regions formed by runaway stars are revealed not to be a negligible source of X-rays and gamma rays in our neighbourhood.

\begin{acknowledgements}

This work was supported by the Spanish Ministerio de Econom\'ia y Competitividad under grant AYA2011-29754-C03-03.

\end{acknowledgements}

\end{document}